\documentclass[a4,aps,pre,twocolumn,10pt]{revtex4-1}
\usepackage[latin1]{inputenc}
\usepackage{amsmath}
\usepackage{amsfonts}
\usepackage{amssymb}
\usepackage{graphicx,hyperref}
\usepackage{amssymb}
\usepackage{epstopdf}
\usepackage{mathptmx}
\usepackage[usenames,dvipsnames]{color}
\newcommand{\be}{\begin {equation}}
\newcommand{\ee}{\end {equation}}
\newcommand{\beqa}{\begin {eqnarray}}
\newcommand{\eeqa}{\end {eqnarray}}
\newcommand{\mb}{\mathbf}
\hypersetup{
    colorlinks,    citecolor=black,    filecolor=black,    linkcolor=black,    urlcolor=black
}

\begin{document}
\title{Magnetic mirror cavities as THz radiation sources and a means of quantifying radiation friction}
 \author{Amol R. Holkundkar}
\email{amol.holkundkar@pilani.bits-pilani.ac.in}
\affiliation{Department of Physics, Birla Institute of Technology and Science, Pilani, Rajasthan, 333031, India}
\author{Chris Harvey}
\email{cnharvey@physics.org}
\affiliation{School of Mathematics and Physics, Queen's University Belfast, Belfast, Northern Ireland, UK}

\begin{abstract}
We propose a radiation source based on a magnetic mirror cavity.  Relativistic electrons are simulated entering the cavity and their trajectories and resulting emission spectra are calculated.  The uniformity of the particle orbits is found to result in a frequency comb in terahertz range, the precise energies of which are tuneable by varying the electron's $\gamma$-factor.  For very high energy particles radiation friction causes the spectral harmonics to broaden and we suggest this as a possible way to verify competing classical equations of motion.
\end{abstract}

\maketitle

\section{Introduction}
It is a well known and extensively utilized result of electrodynamics that an accelerated charge will radiate~\cite{jackson}.  By controlling the particle acceleration with carefully chosen configurations of the background electromagnetic field, light sources can be tailored to specific requirements.  This is the principle behind many devices, such as free-electron lasers, which use undulators to force an electron to oscillate in a controlled way, resulting in the emission light of exact frequencies~\cite{FEL}.  It is also the fundamental precept governing more complex radiation sources such as higher harmonic generation from laser-atom interactions.  Similarly, by firing an electron through a magnetic cavity one can create a gyrotron~\cite{1979dsn..nasa....8K}, a type of free electron maser used as a millimeter wave heat source which has many applications in industry as well as being used to heat plasmas for nuclear fusion experiments.  
In particular, the design and  development of terahertz (THz) radiation sources is a field of contemporary interest because of the vast variety of applications it promises. THz radiation sources are useful in such areas as real time imaging \cite{lee_imaging}, biological research \cite{eric_dna,Alexandrov20101214} and security, etc. In view of this there are various devices are being developed including terahertz semiconductor-heterostructure lasers \cite{kohler2002terahertz}, quantum cascade lasers \cite{cascade_laser,Fathololoumi:12}, etc. 

Here we present (to the best of our knowledge) a previously unconsidered setup in this context -- a magnetic cavity mirror trap.  The device consists of a strong magnetic field with a `basin'-shaped/inverted super-Gaussian profile.  The field strength increases steeply at the ends of the cavity so that particles injected into it will be reflected back towards the central region if they migrate too closely to the edges.  As we shall see, the result is that electrons injected into the cavity will gyrate in a highly regular manner, producing radiation of a very narrow bandwidth.  Although the particles will also acquire a longitudinal drift, they will remain in the cavity due to reflection, bouncing back and forth off each end of the trap.  The primary aim is to generate a tuneable, high quality THz radiation source that can be utilized in many of the applications described above.  Additionally, we will find that such a setup could potentially serve as a testbed for classical theories of radiation friction (RF).

\begin{figure}[b]
\centering \includegraphics[totalheight=3in,trim=1cm 1cm 1cm 1cm]{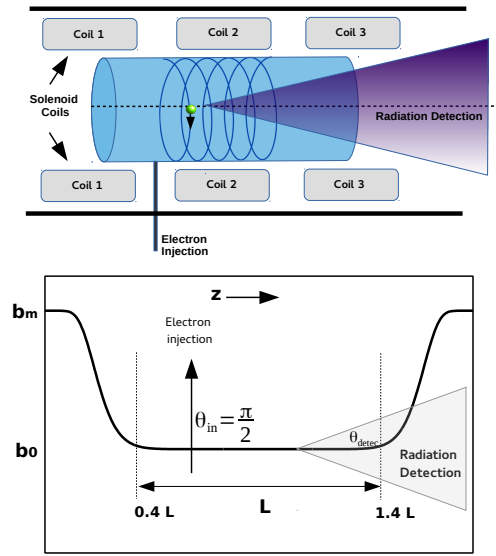}
\caption{A schematic showing the possible experimental setup (top) along with the geometry of the simulation (bottom). $L$ is the length of the cavity, $b_0$ and $b_m$ represent the magnetic field strength in the center region and at the ends, respectively. The radiation is detected
along the $z$ axis.}
\label{geometry}
\end{figure}

\section{Simulation Model}
We consider the  propagation of a relativistic electron in a magnetic field. The spatial profile of the
magnetic field is given by,
\be
B_z(z) = B_{m} - (B_{m} - B_{0})\ \exp\left[-\left(\frac{z - 0.9 L}{0.7 L}\right)^{12}\right].
\label{b1}
\ee
Here, $B_0$ and $B_m$ are the minimum and maximum magnetic field strengths, respectively, and $L$ is
the length of the cavity. The magnetic field has a spatial profile closely resembling an inverted supergaussian.  The field profile and the geometry of our simulation along with the possible experimental setup is illustrated in Fig.~\ref{geometry}.  The desired magnetic field profile can be constructed by placing a set of solenoids around the cavity. The current through them can be calculated using the Biot-Savart law such that it will produce the desired magnetic field as given by Eq.~(\ref{b1}) \citep{garrett}. The radiation from the injected particles is detected along the $z$ axis with $\theta_{detec}$ being the angle of detection.  It should be noted that Gauss's law for magnetism demands that $ \mathbf{\nabla} \cdot  \mathbf{B} = 0$ and so, as a result, the field will also be non-zero along the $x$ and $y$ directions.  Provided ${\partial B_z}/{\partial z}$ does not vary much along the radius of the cavity \cite{chen}, the transverse field components will be,
\be B_x(x,z) = -\frac{x}{2} \frac{\partial B_z}{\partial z}\label{b2},\ee
\be B_y(y,z) = -\frac{y}{2} \frac{\partial B_z}{\partial z}\label{b3}.\ee

The dynamics of a charged particle in a magnetic field is governed by the Lorentz force equations,
\begin{eqnarray}
\frac{d\mathbf{p}}{dt} &=&q[\mathbf{v}\times \mathbf{B}], \label{LF}\\
\mathbf{v} &=&\frac{\mathbf{p}/m_{0}}{\gamma},
\\
\frac{d\mathbf{r}}{dt} &=&\mathbf{v},
\end{eqnarray}
where, $\mathbf{p}$, $\mathbf{v}$, $\mathbf{r}$, $q$ and $m_{0}$ are relativistic momentum, velocity, coordinate, charge and mass of the particle, respectively. Furthermore, $\gamma$ is the relativistic factor and $\mathbf{B}$ is the applied magnetic field. The above equations are solved numerically using a standard Boris leapfrog scheme \cite{filippychev}.  The particle orbit is calculated  by substituting Eqs.~(\ref{b1}) - (\ref{b3}) into the equations of motion, and specifying the initial conditions for the injection energy and the injection angle $\theta_{in}$ (which the trajectory will make with $z$ axis), then the dynamics of the particle are evolved numerically.

Throughout the rest of this article we will work in dimensionless units defined in terms of typical time and length scales of the problem.  Ideally one would normalize in terms of the gyration frequency of the electron, but in our case this quantity is dependent on the initial $\gamma$-factor of the particle.  Instead we normalize in terms of a fundamental frequency $\omega_0=2\pi c/\lambda_0$, corresponding to a wavelength $\lambda_0$ of 1 meter.  We can then define a normalized magnetic field amplitude  $b\equiv qB/m_{e}\omega_0$, where $B$ is the magnitude of the magnetic field in SI units.  
Thus $b_0$ and $b_m$ are the dimensionless magnetic field strengths corresponding to $B_0$ and $B_m$ in Eq. (\ref{b1}-\ref{b3}).
(This is analogous to the dimensionless measure of intensity, $a_0=eE/\omega_0 mc$, used in studies of laser-matter interactions, see for example \cite{Heinzl:2008rh}.)  In these units the dimensionless magnetic field $b = 1$ corresponds to a magnetic field strength of 0.01071 Tesla, the dimensionless unit length, $k_0 x=1$, corresponds to $1/2\pi$ meters and the dimensionless unit time, $\omega_0 t=1$, corresponds to 0.53 ns.  Unless otherwise stated, throughout the rest of this article we will take the cavity length $L$ to be  $\pi/5$ i.e. 0.1 meters.  The particle is injected at an angle $\theta_{in}$ to the field axis  and the subsequent radiation emitted by the particle is detected along the cavity axis ($\theta_{detec} = 0$), as shown in Fig.~\ref{geometry}. In next section we consider specific parameter values.

Next we consider the radiation emitted by the accelerated charge.  The energy radiated per unit solid angle per unit frequency is given by \cite{jackson},
 \be 
\frac{d^2I}{d\omega\ d\Omega} 
 = \frac{e^2}{4\pi^2 c} \left|\int\limits_{-\infty}^{\infty}\frac{\mb{n}\times[(\mb{n}-{\beta})\times\dot{\mb{\beta}}]}{(1 - \beta\cdot \mb{n})^2}       
 e^{i \omega [t' + R(t')/c]} dt' \right|^2,
\ee
where $\mb{n}$ is a unit vector pointing from the particle's position to the detector ($R$) located far away along the $z$-axis, and $\beta$ and $\dot{\beta}$ are, respectively, the particle's velocity and acceleration. In our dimensionless units, if $s = \omega/\omega_0$ is taken to be the harmonic of fundamental frequency, then above equation simplifies to
\be 
\frac{d^2I}{d\omega\ d\Omega} 
 = \left|\int\limits_{-\infty}^{\infty}\frac{\mb{n}\times[(\mb{n}-{\beta})\times\dot{\mb{\beta}}]}{(1 - \beta\cdot \mb{n})^2}       
 e^{i\ s [\tau' + R(\tau')]} d\tau' \right|^2. 
\label{specExp}
\ee
Here $\tau\equiv\omega_0 t$ and we have normalized the intensity by the factor $e^2/(4\pi^2 c)$. All the quantities in the above equations are evaluated at the retarded time so one can directly do the integration in some finite limit. 

The Larmor radius for a relativistic particle in dimensionless units is given by $r_L = p_{\perp}/b_0$, where $p_{\perp}$ is the perpendicular component of the momentum. The cyclotron frequency for a relativistic electron is given by $\omega_c = e B / (\gamma\ m_e)$ which in our dimensionless parameters can be written as $\omega_c' = \omega_0 b_0/\gamma$. 
Finally, the spectrum in units of the cyclotron frequency can be obtained via $\omega = n_0 \omega_c'$, which implies $\omega/\omega_0  = n_0 b_0/\gamma$ and so we need to simply divide by $b_0/\gamma$ to have a spectrum in harmonics of the cyclotron frequency (here $n_0$ is the harmonic number). 

\section{Geometry of the problem}

\begin{figure}[b]
\centering \includegraphics[totalheight=3.5in]{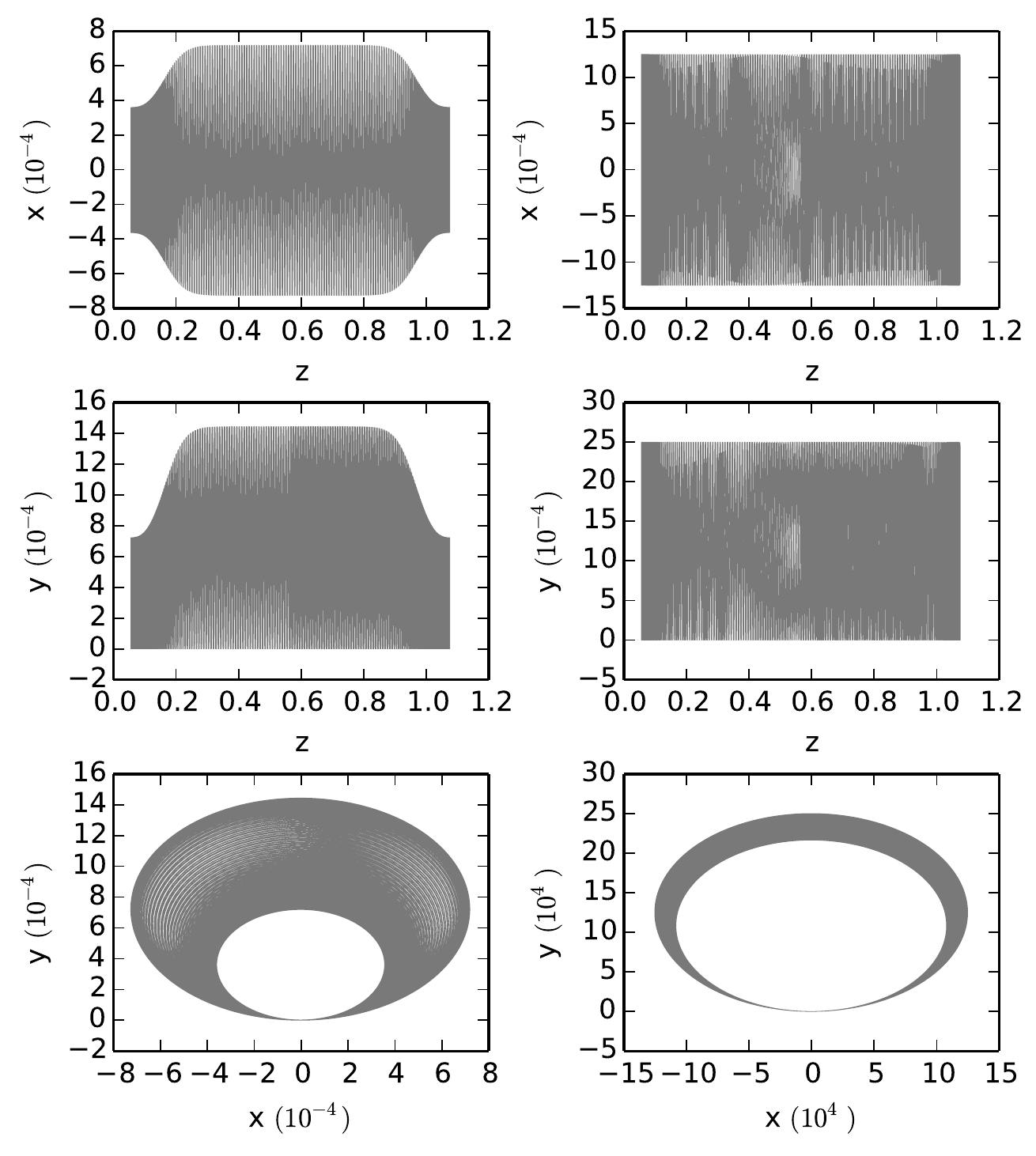}
\caption{The trajectories of the particle for injection angles of $\pi/6$ (left column) and $\pi/3$ (right column).  The orbits are too closely spaced to be individually resolved.}
\label{trajectory}
\end{figure}

The physics of magnetic mirrors is quite well understood and routinely used in plasma confinement machines such as tokomaks. The basis behind the working of magnetic mirrors, and thus also plasma confinement, is the invariance of the magnetic moment.  As the particle moves from a weak magnetic field region to a strong field region the perpendicular component of its velocity ($v_\perp$) must increase to keep the magnetic moment constant. Since the total energy must remain constant, the parallel component ($v_\parallel$) of the velocity must necessarily decrease \cite{chen}. If the field at the end of the cavity is strong enough then eventually $v_\parallel$ becomes zero and the particle is reflected back. The pitch angle with which the particle is injected in the low field region also plays very crucial role in determining the trapping of the particle in the magnetic mirror. 

In order to gain some insight on the dynamics of the particle in this type of magnetic field setup, we have presented the time evolution of the particle trajectory in Fig.~\ref{trajectory}. We have considered the cases where a 511 keV electron ($\gamma_0$ = 2) is injected into the cavity at angles $\theta_{in} = \pi/6$ and $\pi/3$. The value for the magnetic field at the center is chosen to be $b_0 = 1200$, however the value at the end is determined by the criteria that the particle should be trapped inside the cavity and for that to happen the magnetic field at the ends should be at least \cite{chen}, 
\be b_m = b_0/\sin^2\theta_{in} \label{mirrorEq}.\ee  
It can be inferred from Eq.~\ref{mirrorEq} that for smaller injection angles the required value of  $b_m$ increases. For $\theta_{in} = \pi/6$ we chose $b_m = 4810$, which is slightly higher than the threshold magnetic field (4800). Similarly for $\theta_{in} = \pi/3$ we chose $b_m = 1610$, which is also slightly higher than the threshold value (1600). The simulation ran for $\tau $ = 10 and the trajectories of the particle for these two cases are presented in Fig. \ref{trajectory}. The left column shows the result for $\theta_{in} = \pi/6$ and the right column for $\theta_{in} = \pi/3$. As can be seen from these plots, as we increase the injection angle the pinching effect at the ends reduces as a consequence of the lower magnetic field (Eq. (\ref{mirrorEq})) and hence the trajectories become more and more uniform. 

Keeping this aspect of the magnetic field setup in mind, we find that it is most efficient to inject the particle perpendicularly to the magnetic field, since in this case the magnetic field strengths at the ends will be comparable to the field at the center of the cavity, as can be inferred from Eq. (\ref{mirrorEq}). We also note that the field strengths along the transverse directions will be zero exactly at the center ($0.9 L$) of the cavity (Fig.~\ref{geometry}), and so in case of perpendicular injection at this point there will no force along the horizontal direction.  This will not affect the properties of the radiation spectrum and, regardless, such a situation is unlikely to be achieved in an experiment due its sensitivity to the initial conditions.  In fact the nature of the magnetic mirror effect is to stabilize the system leading to a robust experimental setup.

\begin{figure}[t]
\centering \includegraphics[totalheight=2.5in]{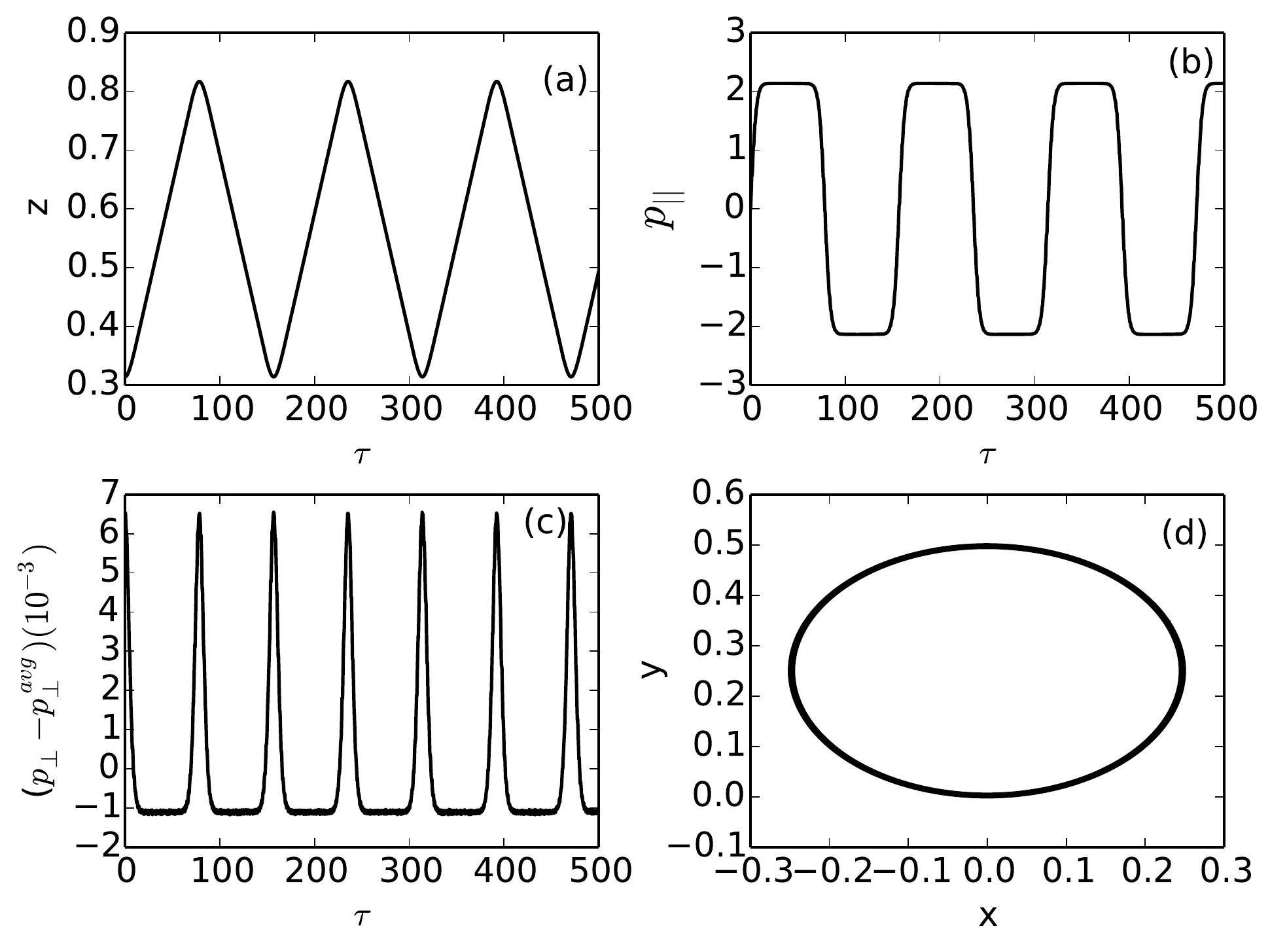}
\caption{The trajectory and momentum of an electron ($\gamma_0 = 300$) injected at an angle $\theta_{in} = \pi/2$ into a cavity with $b_0 = 1200$, $b_m = 1250$.  The plots show the temporal evolution of the $z$ coordinate (a), the parallel component of the momentum $p_\|$ (b), the perpendicular component of the momentum $p_\bot$ (where $p_\bot^{avg} = $ 299.991834) (c)  and the trajectory in the $x-y$ plane (d). }
\label{trajec90}
\end{figure}

The trajectory of a 153 MeV ($\gamma_0 = 300$) particle injected at an angle $\theta_{in}= \pi/2$ at $0.5 L$ is presented in Fig.~\ref{trajec90}. In this case we have chosen our parameters such that the central magnetic field is $b_0 = 1200$ and the magnetic field at the ends is $b_m = 1250$.  The time evolution of the $z$ coordinate of the particle is presented in Fig.~\ref{trajec90}(a), where it can be seen that the particle keeps bouncing between the two ends of the tube. The magnetic field at the ends ($b_m$) influences the frequency of this oscillation. Figure~\ref{trajec90}(b) shows the temporal evolution of the parallel component of momentum ($p_\|$), which is directed along the $z$ direction. The momentum stays constant during the transit of the particle in the tube. However, when it reaches an end, the particle is reflected back from the edge of the cavity and the momentum reverses its sign. The temporal evolution of the perpendicular component of the momentum ($p_\bot^2 = p_x^2 + p_y^2$ ) is presented in Fig.~\ref{trajec90}(c).  Since there is only a slight difference in magnetic field strength between the center and ends of the cavity, the variation in $p_\bot$ is very small.  Therefore we have plotted the deviation from the arithmetic mean as a function of time. The $p_\bot$ increases slightly as the particle reaches the end, because the magnetic field is higher compared to the field in the center. The trajectory of the particle in the $x-y$ plane is also shown in Fig.~\ref{trajec90}(d), where it can be seen that the orbits are very stable in this plane, with a gyration radius given by $r_L = p_\bot/b_0 \approx 300/1200 = 0.25$. The magnetic mirror setup can be seen to be mimicking the dynamics of a particle in a very long cavity with a uniform magnetic field. 

\section{Results and Discussion}

\begin{figure}[t]
\centering \includegraphics[totalheight=2.5in]{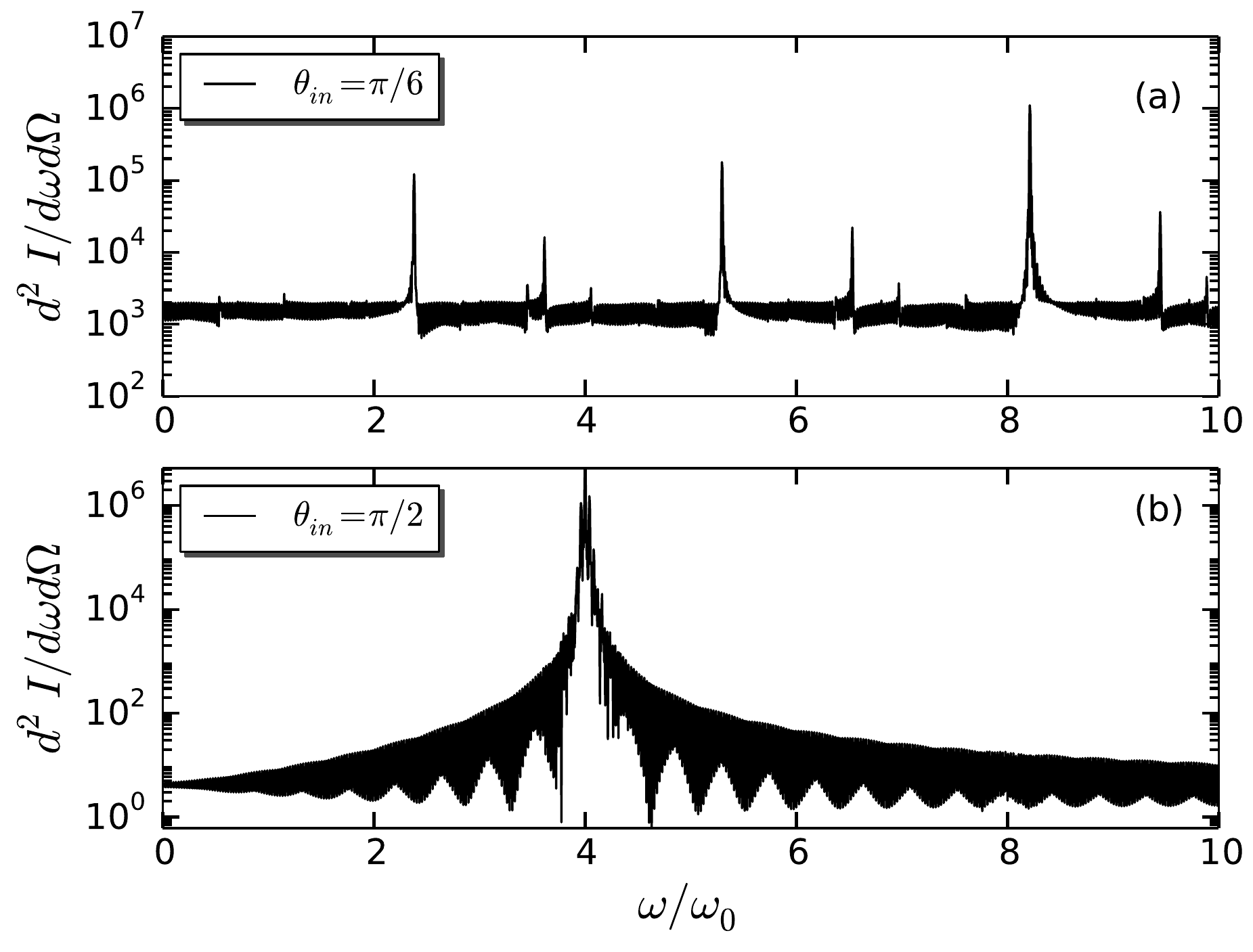}
\caption{The radiation spectrum detected along the cavity for the case when 153 MeV ($\gamma_0 = 300$) electron is injected at $\pi/6$ (a) and at $\pi/2$ (b) to the cavity.}
\label{spec3090}
\end{figure}

In this section we will study the effects of the electron energy and the magnetic field strength on the properties of the emitted radiation.  Before proceeding we first stop to reconsider the influence of the injection angle on the particle dynamics.  In the previous section we saw how perpendicular injection is essential if the particle is to maintain a very stable and uniform orbit.  It would seem likely that the more uniform the orbit the cleaner the corresponding radiation spectrum will be.  In Fig.~\ref{spec3090} we test this by plotting the spectra for the cases when a 153 MeV ($\gamma_0 = 300$) electron is injected in a cavity having $b_0 = 1200$, $b_m = 4850$ and $1250$  for $\theta_{in} = \pi/6$   and $\pi/2$, respectively.
It can be seen that, for the case of  $\theta_{in} = \pi/2$ (Fig.~\ref{spec3090}(b)), the spectrum is indeed very clean, peaking at exactly the cyclotron frequency ($b_0/\gamma_0$) corresponding to the central magnetic field.  On the other hand, the spectrum for $\theta_{in} = \pi/6$ (Fig.~\ref{spec3090}(a)) is a lot messier because there are a range of frequency components contributing due to the varying orbit radius at the ends of the cavity.  Since we are interested in utilizing this setup as a radiation source we will focus our attention on perpendicular injection rather than oblique.

The oscillatory nature of the orbit makes it likely that there will be some similarities with the spectrum produced by an electron in a plane wave field (see, for instance, \cite{PhysRevD.1.2738,PhysRevA.79.063407}).  However, despite similarities, the two cases are not equivalent.  For example, the invariants $B^2-E^2$, $\mathbf{B}\cdot\mathbf{E}$ of the the two fields are not the same and it is never possible to boost to a frame where a plane wave becomes a constant $B$ field.  Nevertheless, we should keep in mind the various plane wave results as we analyze the spectra.  In particular, the high degree of periodicity of the particle orbits will likely result in spectral peaks of a very narrow bandwidth \cite{Harvey:2012ie}, and this is indeed consistent with what we've already seen in Fig.~\ref{spec3090}(b).
 
\begin{figure}
\includegraphics[width=1.0\columnwidth,clip=true,viewport=40 210 540 590]{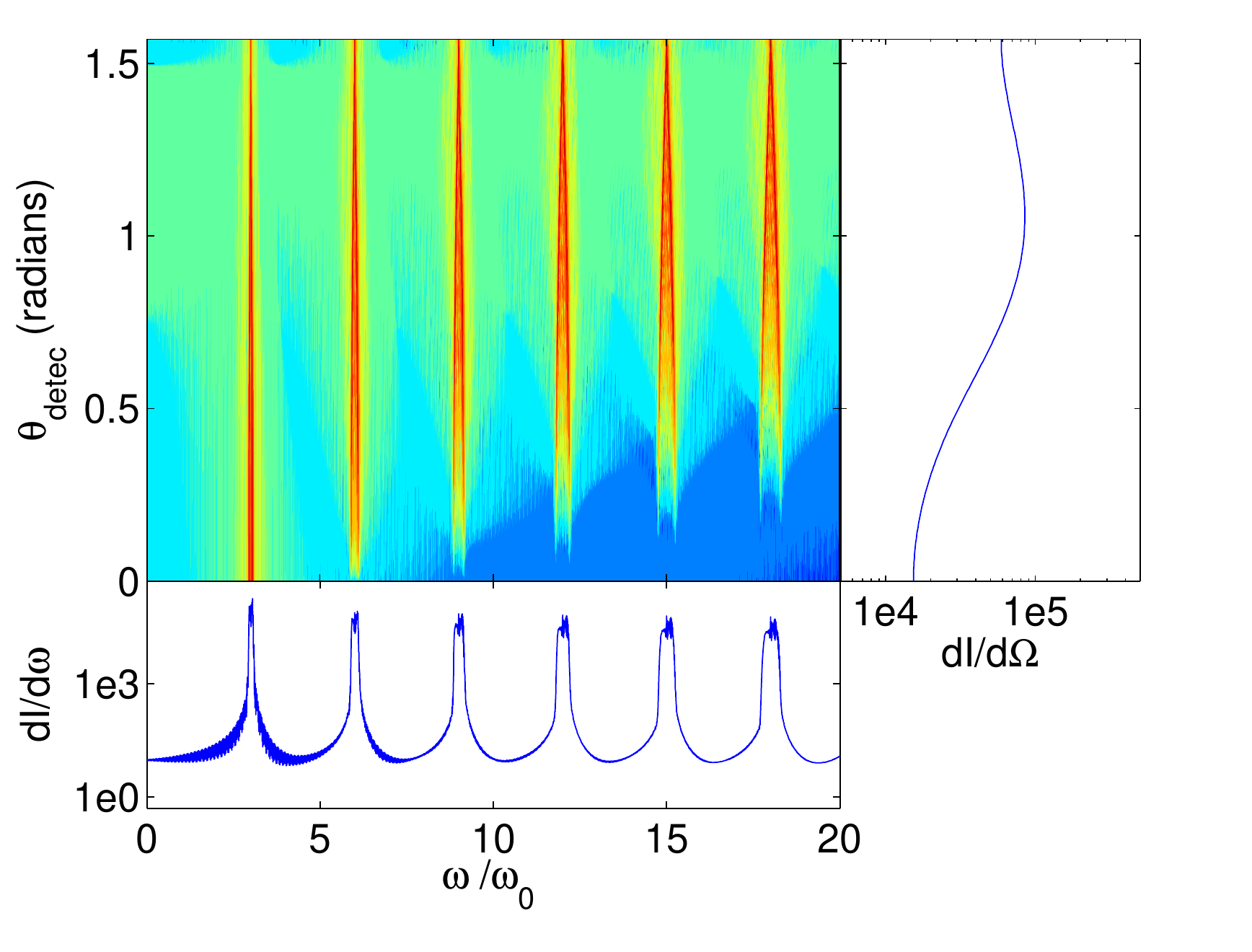}
\caption{The emission spectrum for a particle of $\gamma_0=4$ in a cavity with $b_0=12$, $b_m=15$.  The center plot shows the intensity of the emitted radiation as a function of the detection angle $\theta_{detec}$ and the dimensionless frequency $\omega/\omega_0$.  The lower plot shows the radiated power as a function of frequency (summed over $\theta_{detec}$) and the right hand plot the radiated power as a function of the angle (summed over frequency).   }
\label{fig:total_g_4_b0_12_bm_15}
\end{figure}
\begin{figure}
\includegraphics[width=1.0\columnwidth,clip=true,viewport=40 210 540 590]{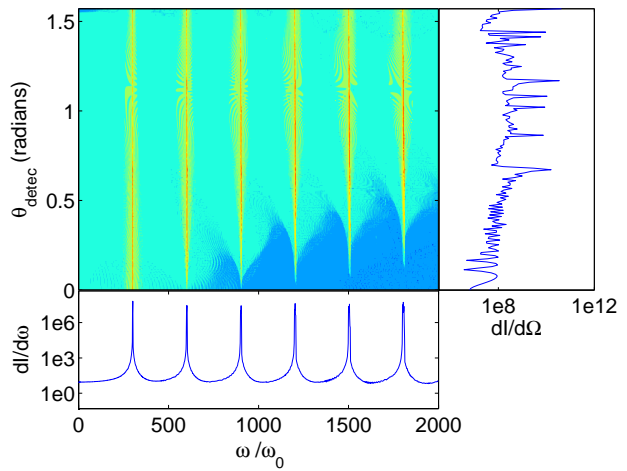}
\caption{The emission spectrum for a particle of $\gamma_0=4$ in a cavity with $b_0=1200$, $b_m=1250$.  The center plot shows the intensity of the emitted radiation as a function of the detection angle $\theta_{detec}$ and the dimensionless frequency $\omega/\omega_0$.  The lower plot shows the radiated power as a function of frequency (summed over $\theta_{detec}$) and the right hand plot the radiated power as a function of the angle (summed over frequency). \label{fig:total_g_4_b0_1200_bm_1250}}
\end{figure}

We begin by considering the radiation emission spectrum for (relatively) low energy electrons, such as those that could be generated using a conventional electron gun.  The spectrum as a function of the emission angle and the radiated frequency is plotted in Figures \ref{fig:total_g_4_b0_12_bm_15} and \ref{fig:total_g_4_b0_1200_bm_1250} for an incoming electron of $\gamma_0=4$, for two different magnetic field strengths.   It can be seen that in both cases the harmonics are very clean, appearing at integer multiples of the respective cyclotron frequencies ($b_0/\gamma_0$).  
In the cases we are considering there will be many harmonics contributing to the spectrum, the total number scaling like $\gamma^3$ \cite{jackson}. Thus the total range of the spectrum can, depending on the parameters, go from the THz to the XUV range.
However, calculating the full spectrum is computationally expensive and so we have just plotted the first few harmonics here for the purposes of our discussion. Observe that the intensity of the radiation is roughly uniform over the whole angular range of each harmonic.  This is because the angular width of the emissions go like $\sim 1/\gamma$ which is of the same order of magnitude as the angular range we are considering \cite{jackson}.  Note also that the fundamental (i.e.~first) harmonic is the only one that contributes precisely along the cavity axis $\theta_{detec}=0$.  This is consistent with the emissions in a plane wave, where the spectrum exhibits a `dead-cone' at $\theta_{detec}=0$ \cite{PhysRevA.79.063407}.  As we move to higher harmonics the direction they contribute in gradually moves away from the cavity axis.

Since $\gamma_0$ is the same in these two cases, we can assess the influence of the magnetic field strength on the properties of the spectrum.  Because the harmonics are at integer multiples of the cyclotron frequency $b_0/\gamma_0$, as we increase the strength of the magnetic field the harmonics will be blue shifted.  (At first sight, this appears to be the opposite to what happens in the case of an electron in a plane wave field, where an increase in the wave amplitude causes the harmonics to be red shifted.  However in this case it is conventional to define the field intensity such that it is normalized by the frequency of the wave.  In our case there is no fundamental frequency due to the background field -- the timescales come from the size of the particle's orbit.  Since this  is dependent on the $\gamma$-factor, we are not comparing like with like.)

\begin{figure}[t]
\centering \includegraphics[totalheight=2.5in]{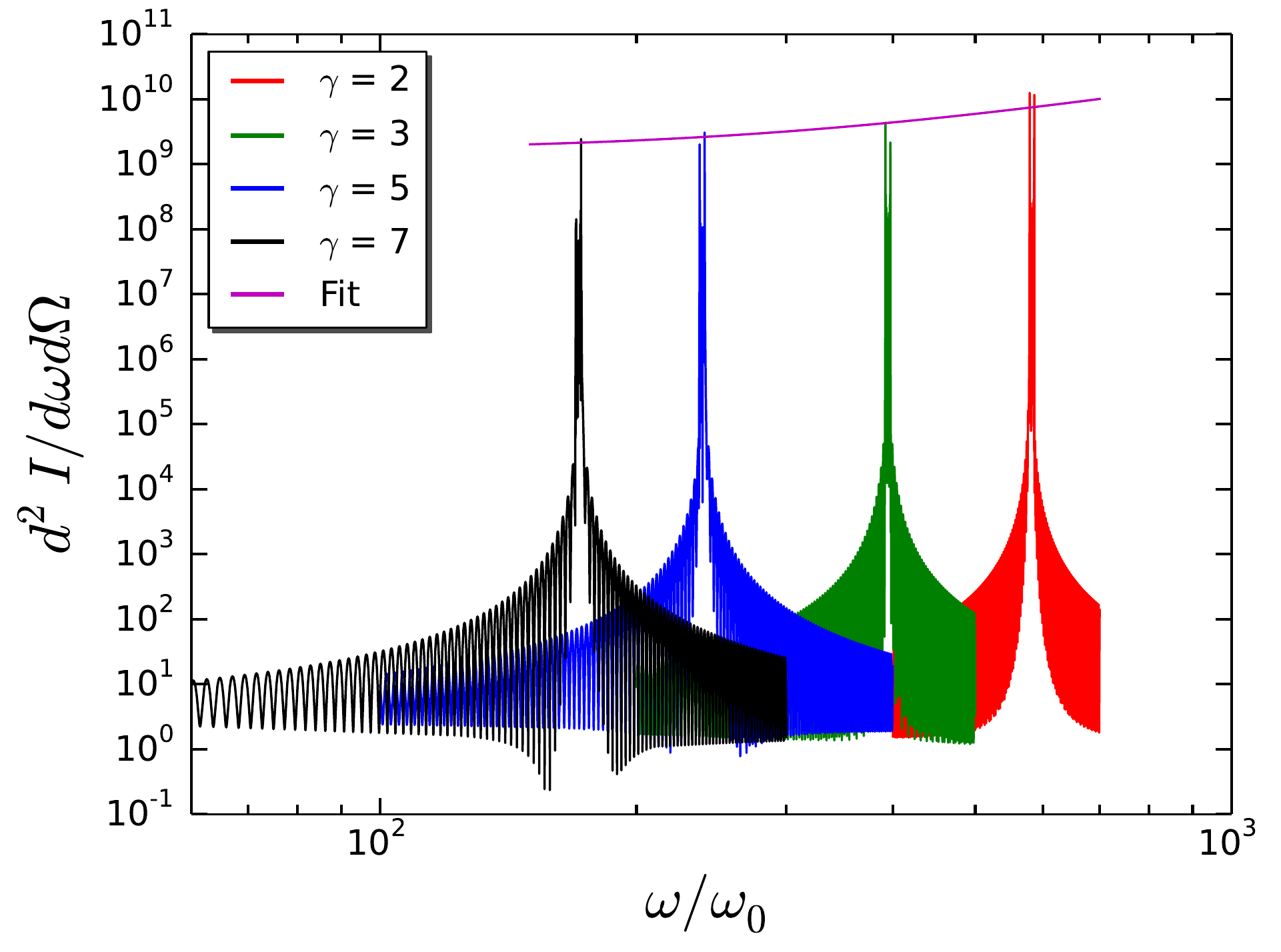}
\caption{The radiation spectrum emitted by the gyrating particle with $b_0 = 1200$ and $b_m = 1250$ for different energies. The fit line shows the variation of the peak amplitude with particle energy. It is observed that the peak amplitude varies as $1/\gamma^2$.  }
\label{speclowGama}
\end{figure}

Next we consider the dependence of the spectra on $\gamma_0$.  Since we have already established that the emission harmonics occur at multiples of the cyclotron frequency we expect that, for magnetic fields of a fixed strength, the frequencies of the spectrum will be redshifted by a factor of $1/\gamma_0$ as we increase the electron energy.  In Fig.~\ref{speclowGama} we investigate this by fixing the fields to be $b_0 = 1200$ and $b_m = 1250$  and plotting the fundamental harmonic in the $\theta_{detec}=0$ direction for $\gamma_0 = 2$, 3, 5 and 7.  As expected, the frequencies of the harmonic peaks decrease as we increase the $\gamma$-factor.

Finally, we will examine how the amplitude of the fundamental harmonic changes with $\gamma_0$.  It is known from the standard textbook theory that a particle undergoing instantaneous circular motion will radiate a power spectrum that has the form
\begin{eqnarray}
dP \sim \frac{1}{\gamma^2}\frac{\beta B}{(1-\beta\sin\theta)^3}\bigg(1-\frac{\cos^2\theta}{\gamma^2(1-\beta\sin\theta)^2}\bigg),\label{dPcirc}
\end{eqnarray}
where we have abbreviated $\theta_{detec}=\theta$ (see, for example, Ch.~14 of Ref.~\cite{jackson}).  Along the cavity axis the radiated power will go like $\beta B(1-1/\gamma^2)/\gamma^2$ and so we can expect a $1/\gamma^2$ scaling of peak emission amplitude with the particle energy.  In Fig.~\ref{speclowGama} we plot a fit-line of the peak amplitude as a function of particle's energy.  It is observed that amplitude indeed decreases like $ 1/\gamma^2$.  

\begin{figure}[t]
\includegraphics[width=1.0\columnwidth,clip=true,viewport=40 210 540 590]{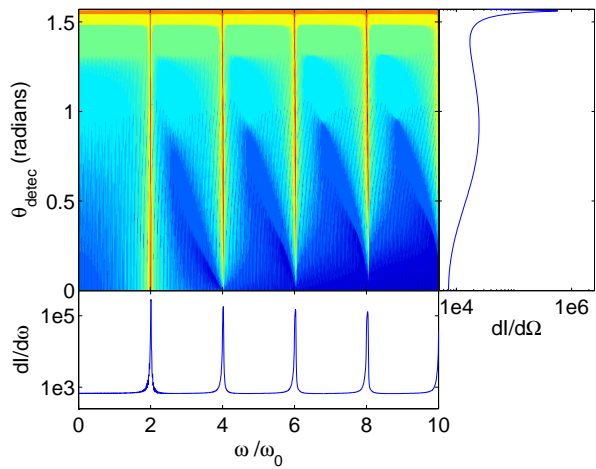}
\caption{The emission spectrum for a particle of $\gamma_0=600$ in a cavity with $b_0=1200$, $b_m=1250$.  The center plot shows the intensity of the emitted radiation as a function of the detection angle $\theta_{detec}$ and the dimensionless frequency $\omega/\omega_0$.  The lower plot shows the radiated power as a function of frequency (summed over $\theta_{detec}$) and the right hand plot the radiated power as a function of the angle (summed over frequency). 
\label{fig:total_g_600_b0_1200_bm_1250} }
\end{figure}

 \begin{figure}
\centering \includegraphics[totalheight=2.5in]{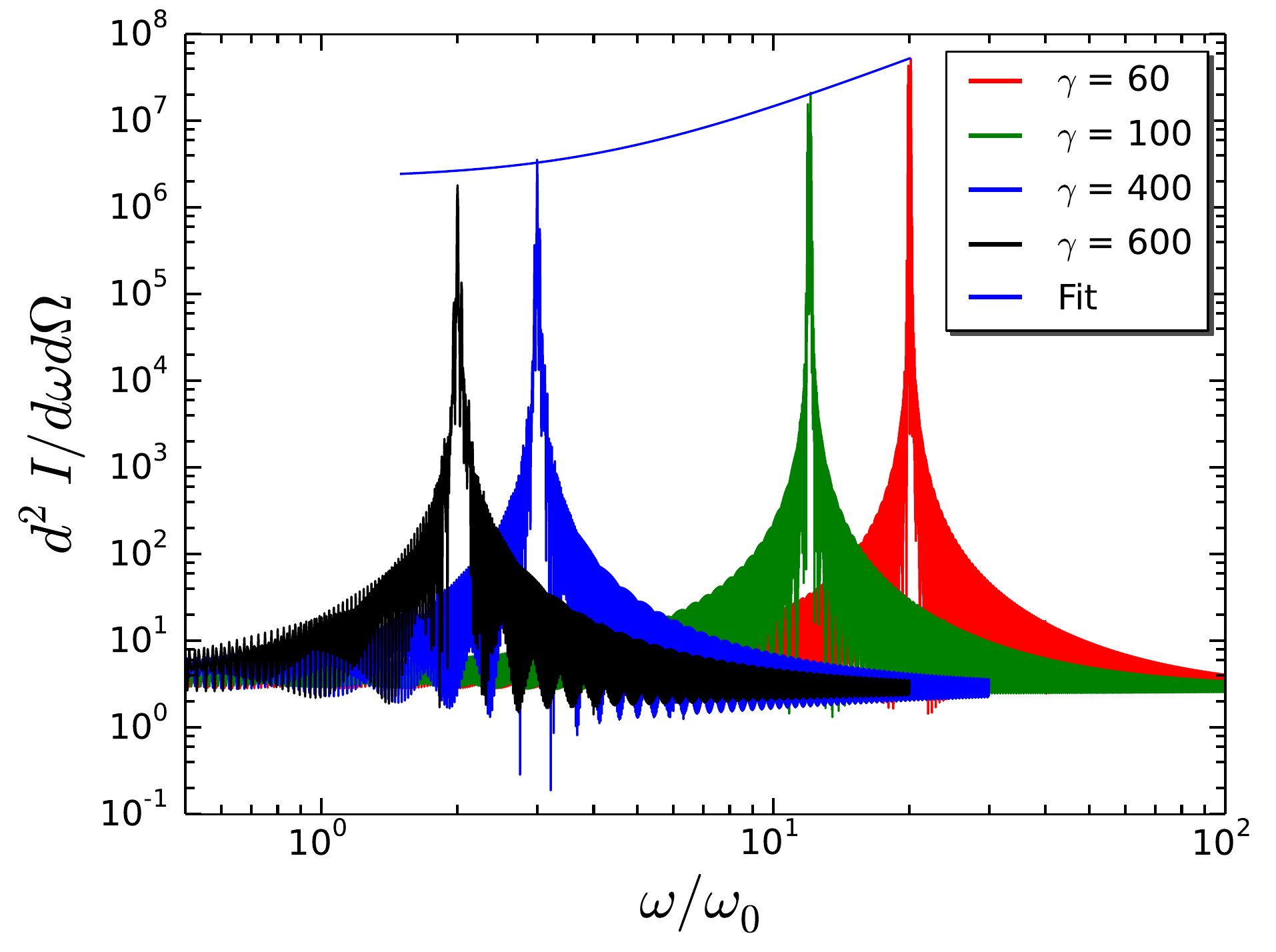}
\caption{The radiation spectrum emitted by the gyrating particle with $b_0 = 1200$ and $b_m = 1250$ for different energies. The fit line shows the variation of the peak amplitude with particle energy. It is observed that the peak amplitude varies as $1/\gamma^2$.   }
\label{specGama}
\end{figure}

We end this section by considering how the system behaves if we switch to an ultra-relativistic electron source, such as from a particle accelerator.  In Fig.~\ref{fig:total_g_600_b0_1200_bm_1250} we show the emission spectrum for an electron with $\gamma_0=600$ inserted into the same cavity as before, $b_0 = 1200$ and $b_m = 1250$.  Once again the harmonics are very clean, occurring at integer multiples of the cyclotron frequency.  The main difference in this case is that there is a strong peak of broadband radiation emitted in the direction perpendicular to the cavity ($\theta_{detec}=\pi/2$).  This is because an ultra-relativistic particle will radiate most of its energy in its direction of motion~\cite{jackson}, which in this case will be tangential to the circular orbit.   We note that the amplitude of the these harmonics are significantly lower than in the previous, lower energy cases.  This is because the radiated power is more concentrated towards the perpendicular direction, as can be seen by considering Eq.~(\ref{dPcirc}) in the limit $\theta_{detec}\to\pi/2$, $\beta\sim 1$.

Despite the peak radiation being perpendicular to the cavity, it can still be interesting to detect the radiation along the cavity axis, since here we have a clean frequency comb, rather than the `washed-out' broadband spectrum seen perpendicular to the axis.  However, the conversion efficiency is not as high in this direction.  In Fig.~\ref{specGama} we plot the the fundamental harmonic, viewed along the cavity axis ($\theta_{detec}=0$), for various large $\gamma_0$'s.  Once again we see the same characteristic behavior, with the peak positions being redshifted according to $1/\gamma_0$ and the signal strength falling like $1/\gamma_0^2$.
As a result of this, the high $\gamma$ setup is not as practicable as a radiation source as the lower energy (and more convenient) electron gun scenario, though there will be higher harmonics extending into the XUV range.

Nevertheless, the high energy setup may be interesting from a theoretical point of view, if we consider the effects of RF on the particle dynamics.  We will investigate this in the following section.

\section{Effect of Radiation Friction}
Since the electron in the cavity is radiating it must also be losing energy.  Over short timescales the amount of radiation loss is typically small compared to the particle's energy and so can ordinarily be neglected.  However, if we consider the behavior of the particle over an extended time period then the (instantaneously) small effect of radiation loss will have more of a cumulative impact.  Additionally, if we increase the strength of the magnetic field, or increase the $\gamma$-factor, then the acceleration acting on the particle may be strong enough for RF effects to start to become important over shorter timescales.  It is therefore worth extending our modeling to include RF for two reasons.  Firstly, so that we can check whether it is valid to neglect RF in the regimes we have been considering and, secondly, to explore the potential for using this setup with different parameters to investigate RF in its own right.

Despite having been studied for over 100 years, finding the correct description of the dynamics of a radiating particle remains one of the most fundamental problems in electrodynamics.  The most common starting point is the Lorentz-Abraham-Dirac equation which is obtained by solving the coupled Lorentz and Maxwell's equations \cite{Lorentz:1905,Abraham:1905,Dirac:1938nz}.  However, this equation is notorious due to its defects such as pre-acceleration and (unphysical) runaway solutions.  One of the most common ways of removing these problems is to adopt the perturbative approximation of Landau and Lifshitz \cite{LLII} so that the equation of motion becomes
\begin{eqnarray}
\frac{d\mathbf{p}}{dt}=\mathbf{f}_\textrm{L}+\mathbf{f}_\textrm{R},
\end{eqnarray}
where $\mathbf{f}_\textrm{L}$ is the Lorentz force (\ref{LF}) and the radiative correction term
\begin{widetext}
\begin{eqnarray}
\mathbf{f}_\textrm{R}=-\left( \frac{4}{3}\pi \frac{r_e}{\lambda_0} \right)\gamma\left[ \mathbf{v}\times \left( \frac{\partial}{\partial t}+\mathbf{v}\cdot\nabla\right)\mathbf{B}\right]+\left( \frac{4}{3}\pi \frac{r_e}{\lambda_0} \right) [\mathbf{v}\times\mathbf{B}\times\mathbf{B}]-\left( \frac{4}{3}\pi \frac{r_e}{\lambda_0} \right) \gamma^2 [(\mathbf{v}\times\mathbf{B})^2]\mathbf{v}, \label{LL}
\end{eqnarray}
\end{widetext}
where $r_e\equiv e^2/mc^2\approx 2.8\times 10^{-15}$m is the classical electron radius.  Equation (\ref{LL}) is valid when the radiative reaction force is much less than the Lorentz force in the instantaneous rest frame of the particle.  We note that there are numerous alternative equations in the literature (see, for example, \cite{Sokolov:2009,O'Connell:2012ee}) and it is still an open problem as to which is the correct formulation.  However, the Landau Lifshitz equation has, along with some others, recently been shown to be consistent with quantum electrodynamics to the order of the fine structure constant $\alpha$ \cite{0038-5670-34-3-A04,Ilderton:2013tb}.  
Despite these controversies no experiment has been conducted to distinguish between the different models.  Some studies have investigated the use of high intensity lasers to probe RF effects (see, for example, \cite{Harvey:2011mp,PhysRevE.89.021201}), but such an approach may not be the most optimum for distinguishing between models, because particles inserted into the beam will radiate away much of their energy before they enter the most intense part of the field \cite{PhysRevE.88.011201,PhysRevE.89.053315}. 

\begin{figure}[t]
\centering \includegraphics[totalheight=2.3in]{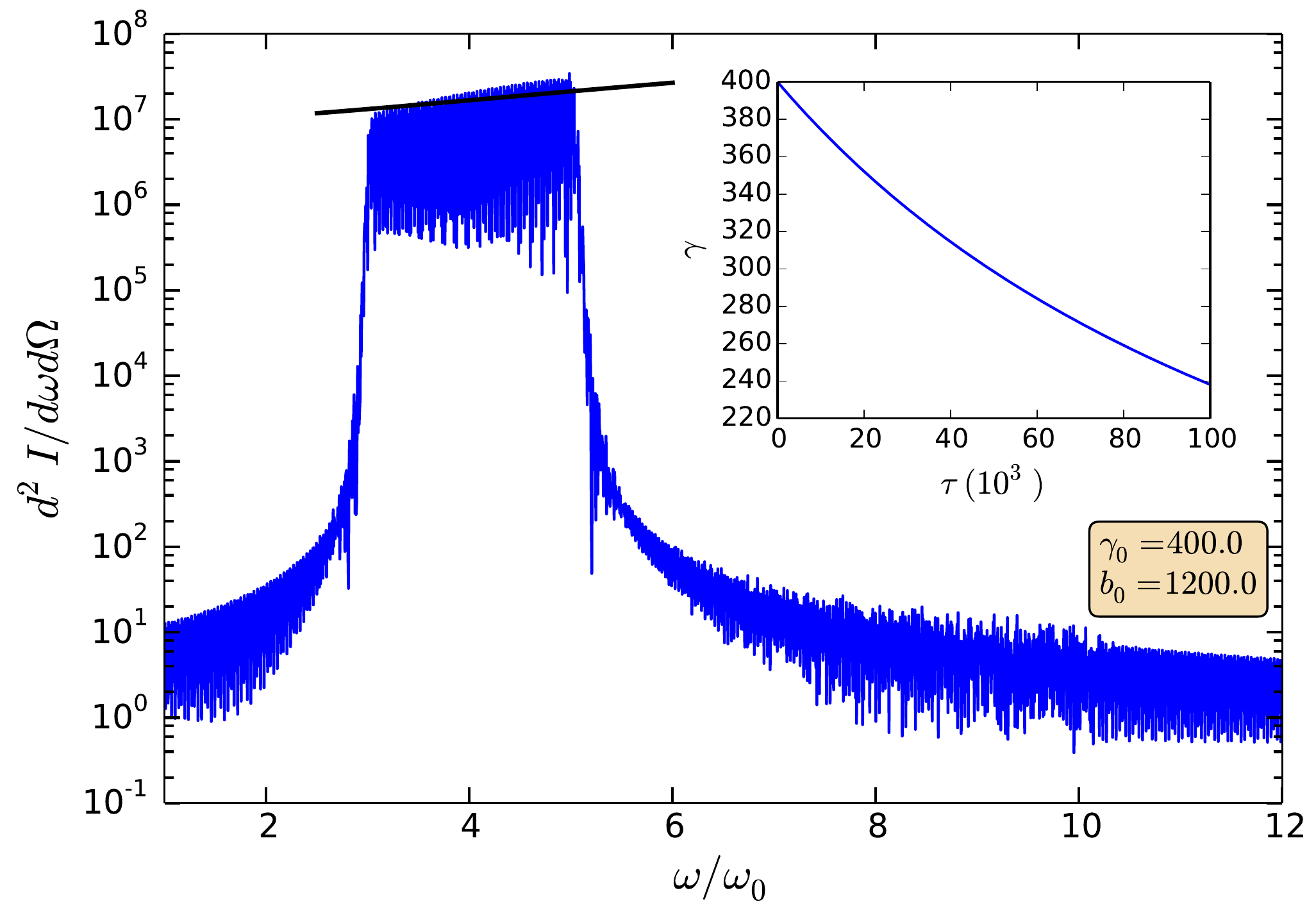}
\caption{Spectrum for the case $\gamma_0$ = 400, $\theta_{in} = \pi/2$, $\theta_{detec} = 0$, $b_0 = 1200$, $b_m = 1250$.  Here RF is included. The fit line again shows a $1/\gamma^2$ scaling.}
\label{spec400RR}
\end{figure}
In Fig.~\ref{spec400RR} we show the spectrum (and, in the inset, the $\gamma$-factor) for the case when $\gamma_0 = 400$ but with RF included.  We see from the inset that the particle gradually looses energy as it moves back and forth along the cavity, falling to about two thirds of its initial value by the end of the simulation ($\tau=1\times 10^5$ is equivalent to 0.053 ms).  From the spectrum we can see that initially the emissions peak at the same value as the case without RF (i.e.~$\omega/\omega_0=3$), as we would expect.  Then, as the particle loses energy, the frequency of the spectral peak gradually increases, resulting in a broadening of the harmonic as compared to the case without RF.  This behavior is consistent with Fig.~\ref{specGama} where we found that particles with a lower $\gamma$-factor radiate at higher frequencies along the cavity axis.  

Thus we can see that, when our high-$\gamma$ magnetic cavity source is used over an extended time period, the frequency harmonics of the emitted radiation are slowly blue shifted as the electron loses energy due to RF.  Although the broadening of the harmonic due to radiation loss is not large, it is nevertheless significant and easily measurable.  This raises the prospect of comparing the spectra calculated using the various competing classical RF theories with the experimental results, potentially allowing us to test the different models.  (This is not completely dissimilar from the idea of cycling an electron through a modified Penning trap, which was proposed in Ref.~\cite{0295-5075-50-3-287}.)  However, such an investigation is beyond the scope of this paper and so we leave it for a future publication.

\section{Concluding Remarks} 
We have presented (to best of our knowledge) a previously unconsidered setup for generating the radiation in THz range. The setup utilizes a magnetic mirror to confine a (relativistic) charged particle to a cavity.  The particle is injected perpendicularly to the applied magnetic field and then the `magnetic-mirror' effect serves to mimic the dynamics of particle in a very long cavity with uniform magnetic field. The result is a uniform, closely-spaced, circular orbit (helical motion) for the particle, causing it to emit a narrow band of radiation at its cyclotron frequency. The emitted radiation will be in the THz range and can be tuned to an exact frequency but changing the particle energy. The simplicity and portability of the proposed 
setup makes it a viable THz radiation source which would be useful for medical research, defense applications, communication purposes, imaging and spectroscopy, etc.
Furthermore, at nonzero detection angles higher harmonics of the fundamental frequency (cyclotron frequency) are also observed.  This promises to yield a frequency comb from the THz to the XUV range (although we have only presented the first few harmonics due to limitations in computational resources).

The effect of RF on the emitted radiation has also been studied. It is observed that the spectrum becomes blueshifted when RF is taken into account. This is because, as the particle looses energy, its cyclotron frequency is reduced by $1/\gamma$ (where here $\gamma$ would be a function of time).  Thus, apart from use as a THz radiation source, the setup could also be exploited to experimentally validate the various competing theories of RF. The extensive analysis of the THz radiation and its harmonics under the different models is beyond the scope of the current manuscript and has been left for a future publication.

\section*{Acknowledgements}
AH acknowledges the Science and Engineering Research Board, Department of Science and Technology, Government of India for funding the project SR/FTP/PS-189/2012.  CH acknowledges support from the EPSRC, UK, Grant No.~EP/I029206/1-YOTTA.

%

\end{document}